\documentclass[pdflatex,sn-aps]{sn-jnl}

\usepackage{graphicx}%
\usepackage{multirow}%
\usepackage{amsmath,amssymb,amsfonts}%
\usepackage{amsthm}%
\usepackage{mathrsfs}%
\usepackage[title]{appendix}%
\usepackage{xcolor}%
\usepackage{textcomp}%
\usepackage{manyfoot}%
\usepackage{booktabs}%
\usepackage{algorithm}%
\usepackage{algorithmicx}%
\usepackage{algpseudocode}%
\usepackage{listings}%
\usepackage{caption}
\usepackage{subcaption}

\begin{document}

\title[Coherent quantum control of dark excitons in hybrid metal-organic chalchogenolates]{Coherent quantum control of dark excitons in hybrid metal organic chalchogenolates}

\author[1,2]{\fnm{Christian L.} \sur{McCoy}}\email{christian.mccoy@uconn.edu}
\equalcont{These authors contributed equally to this work.}

\author[2]{\fnm{Tobias} \sur{Saule}}\email{tobias.saule@uconn.edu}
\equalcont{These authors contributed equally to this work.}

\author[1,3]{\fnm{Mariya} \sur{Aleksich}}\email{mariya.aleksich@uconn.edu}

\author[1,3]{\fnm{Maggie C.} \sur{Willson}}\email{maggie.ward@uconn.edu}

\author[1,3]{\fnm{J. Nathan} \sur{Hohman}}\email{james.hohman@uconn.edu}

\author[4]{\fnm{Thomas} \sur{Weinacht}}\email{thomas.weinacht@stonybrook.edu}

\author[2]{\fnm{George N.} \sur{Gibson}}\email{george.gibson@uconn.edu}

\author[1,2]{\fnm{Carlos A.} \sur{Trallero-Herrero}}\email{carlos.trallero@uconn.edu}

\affil[1]{\orgdiv{Department of Chemistry}, \orgname{University of Connecticut}, \orgaddress{\city{Storrs}, \postcode{06269}, \state{Connecticut}, \country{United States}}}

\affil[2]{\orgdiv{Department of Physics}, \orgname{University of Connecticut}, \orgaddress{\city{Storrs}, \postcode{06269}, \state{Connecticut}, \country{United States}}}

\affil[3]{\orgdiv{Institute of Material Science}, \orgname{University of Connecticut}, \orgaddress{\city{Storrs}, \postcode{06269}, \state{Connecticut}, \country{United States}}}

\affil[4]{\orgdiv{Department of Physics}, \orgname{Stony Brook University}, \orgaddress{\city{Stony Brook}, \postcode{11794}, \state{New York}, \country{United States}}}

\abstract{
Artificial atom-like systems are a promising candidate for next generation quantum processing. Among them, dark excitons exhibit one of the longest lifetimes at high temperatures. Here, we demonstrate coherent control of dark excitonic states in metal-organic chalcogenolates (MOChas) by using an ultrafast pulse shaper at room temperature. These dark exciton states are optically accessed via two-photon absorption and directly read out with a
four-wave mixing process. The system is described by a non-perturbative, two-photon
Hamiltonian based on well-known atomic physics and applied to a three level system
comprised of two dark excitons. Empirical and theoretical state specific optical access is shown via a simple optical pulse shape. The developed Hamiltonian-based description is a first step towards a quantum processing platform using three-level systems and two photon transitions, one example being dark excitons in the MOCha silver benzeneselenolate (mithrene). Simple conditions for gate operations are laid out and described.}


\maketitle

\section*{Introduction}

Coherent quantum control of molecular systems is a very mature field that dates back to the pioneer work of Abraham and Lemoine, Zewail, Warren, Rabitz, Shapiro and Brumer, and many others. Quantum control in condensed media includes processes such as 2D spectroscopy~\cite{cundiffRabiFloppingSemiconductors1994,Cundiff2014,modalLongLivedPolaritonicCoherenece,thomasRecentAdvancesMultidim2018} and THz control~\cite{woernerUltrafastTwodimensionalTerahertz2013,leinssTerahertzCoherentControl2008,borschLightwaveElectronicsCondensed2023}.

One main scientific field driving advances in quantum control is quantum information science or sensing, which requires operations to be completed within the coherence time of its qubits. One solution is using ultrafast lasers that act-upon or measure the quantum state within a femtosecond time scale, well below any decoherence times. These techniques have been used to control natural~\cite{meshulachCoherentQuantumControl1998,meshulachCoherentQuantumControl1999,dudovichTransformLimitedPulsesAre2001,trallero-herreroCoherentControlStrong2005} and ``artificial atoms'', i.e. quantum dots~\cite{kappe2025chirped}. One natural progression is quantum control of naturally-occurring and scalable systems that resemble artificial atoms, such as excitons. 
Excitons are quasi-particles formed from the promotion of an electron from the valence band to the conduction band and its interaction with the remaining electron-hole in the valence band through coulombic interaction. The quasi-particle itself is akin to a synthetic atom in that excitons have discrete energetic states. The parity of these states is an essential selection rule for governing optical transitions. Specifically, in direct band gap semiconductors with interband transitions that are dipole-allowed, single-photon absorption exclusively accesses even-parity exciton states, whereas two-photon absorption (TPA) exclusively accesses odd-parity exciton transitions~\cite{haugQuantumTheoryOptical2004,yeProbingExcitonicDark2014,foxOpticalPropertiesSolids2012} (see Fig.~\ref{fig:transition}(a.)). These states accessible only through TPA are known as exciton dark states or simply dark excitons, due to their lack of appearance in a linear absorption spectrum and lack of direct light emission, making their optical study more challenging~\cite{wangOpticalResonancesCarbon2005,yeProbingExcitonicDark2014,naProbingDarkSide2020,bangeUltrafastDynamicsBright2023,wangUltrafastManybodyBright2023}.

As of late, a new class of material has emerged:
Metal-Organic Chalcogenolate, or MOChas. 
MOChas in their base unit are comprised of an inorganic layer, consisting of a transition metal and a chalcogen atom, and two buffering organic ligand layers. This results in 1-dimensional (1-D) or 2-dimensional (2-D) metal-chalcogen polymer sheets, an example structure is shown in the inset of Fig. \ref{fig:experiment_buildup}.
Due to the two-dimensional confinement nature of MOChas, quantum well properties of the layered inorganic-organic layers~\cite{yaoStronglyQuantumConfinedBlueEmitting2021} emerge. These materials have also been shown to have strong light-matter coupling, with mithrene (silver benzeneselenolate, $\mathrm{[AgSePh]_{\infty}}$) being one of the strongest exciton-producing materials~\cite{schriberMithreneSelfAssemblingRobustly2018,maseratiAnisotropic2DExcitons2021,yaoStronglyQuantumConfinedBlueEmitting2021,anantharamanUltrastrongLightMatter2025}. The strong exciton binding energy of mithrene promises exciting quantum sciences applications at near room temperature. In addition, these materials are very economical, easy to manufacture, easy to modify, and very stable~\cite{yeungCorrosionLateandPostTransition2020}.

In this paper, we present experimental as well as theoretical access to the dark exciton states in mithrene using an ultrafast pulse shaper and a modified three-level Hamiltonian for modeling the interactions. We show that we not only have access to the populations of the $\mathrm{B_{2p}}$ state, but we perform first proof-of-concept coherent control experiments using a tailored spectral phase to target the population of the $\mathrm{B_{2p}}$ exciton state. Knowing the Hamiltonian of such a system will allow for the design of quantum gates.

\section*{Results and Discussion}

\subsection*{Dark Excitons Beyond the Perturbative Limit}

Dark excitons are formed after the resonant absorption of two photons. From a perturbative perspective, the third order susceptibility produces a very sensitive probe of the state of the quasiparticles at frequencies that are discernible with respect to the decay frequency of the bright exciton.

The lowest order non-linear optical process in a centrosymmetric ($\mathrm{\chi^{(2)}=0}$) system will be dominated by $\mathrm{\chi^{(3)}}$ which governs the coupling between all states. In our case, the third order susceptibility of a two photon absorption followed by an emission to the ground state is proportional to the following~\cite{boydNonlinearOptics2020}:

\begin{equation}
\mathrm{\scalebox{1}{$
       \chi^{\left(3\right)}_{kjih} \left(\omega_{\mathrm{FWM}}, -\omega_{2}, \omega_{1}, \omega_{1} \right)
        \propto
        \sum_{mn\nu}\frac{\boldsymbol{\mu}^j_{g\nu}\boldsymbol{\mu}^k_{\nu n}\boldsymbol{\mu}^i_{nm}\boldsymbol{\mu}^h_{mg}}{\left(\omega^{\ast}_{\nu g} - \omega_{2}\right)\left(\omega_{\mathrm{A_{2p} /B_{2p}}}-2\omega_{1}\right)\left(\omega_{mg}-\omega_{1}\right)}
               $}
}\label{eq:Chi_3_case} 
\end{equation}

\begin{figure}
    \centering
    \includegraphics[width=0.75\linewidth]{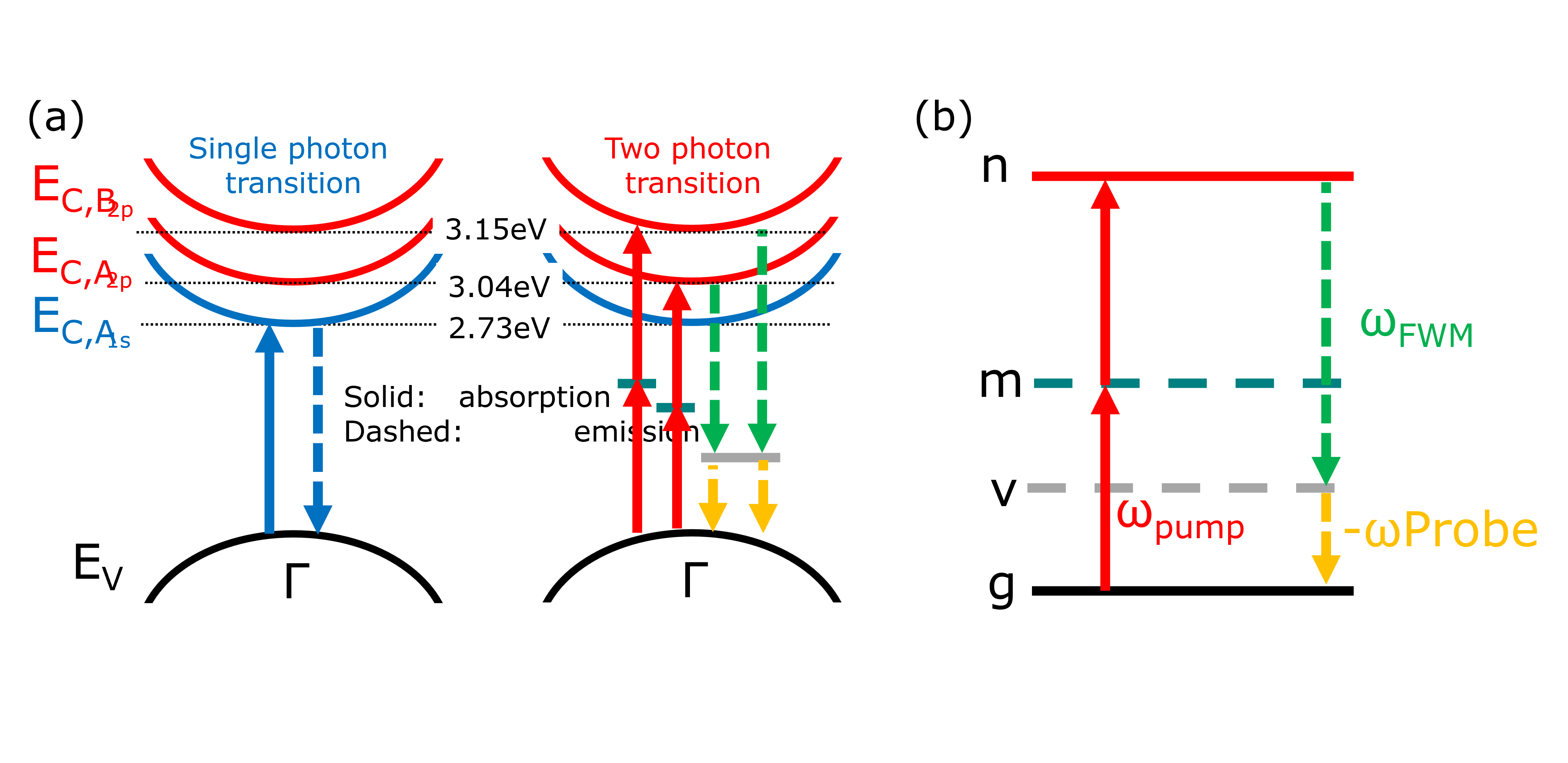}
    \caption{\textbf{(a.)} Excitonic state representation of mithrene, single photon allowed transitions in blue (left). Two photon allowed transitions in red (right). \textbf{(b.)}  State diagram of a $\mathrm{\chi^{\left(3\right)}}$ process. Formalism and nomenclature from Eq.~\ref{eq:Chi_3_case} and Ref.~\cite{boydNonlinearOptics2020}. In the case of mithrene the only real n-states are the dark exciton states, $\mathrm{A_{2p}}$ and $\mathrm{B_{2p}}$.}
    \label{fig:transition}
\end{figure}

\vspace{3mm}

with $\mathrm{\omega_{FWM}}$, being the observed emission (FWM, green in Fig.~\ref{fig:transition}), $\mathrm{\omega_{1}}$, the pump beam (red in Fig.~\ref{fig:transition}), and $\mathrm{\omega_{2}}$ the probe beam (yellow in Fig.~\ref{fig:transition}). Those, Eq.~\ref{eq:Chi_3_case} can be also represented as $\mathrm{\chi^{(3)}(\omega_{\mathrm{FWM}},\omega_2, \omega_1, \omega_1) = \chi^{(3)}(\omega_{\mathrm{FWM}}, \omega_{\mathrm{probe}}, \omega_{\mathrm{pump}}, \omega_{\mathrm{pump}})}$. $\mathrm{\boldsymbol{\mu}^j_{g\nu},\boldsymbol{\mu}^k_{\nu n},\boldsymbol{\mu}^i_{nm},\boldsymbol{\mu}^h_{mg}}$ are the couplings or dipole moments between the virtual states $\mathrm{\nu,m}$ and real states $\mathrm{g,n}$. With $\omega_{mg}$ and $\omega^{\ast}_{\nu g}$ being virtual states and $\mathrm{\omega_{A_{2p} /B_{2p}}}$ being two independent, real states. A schematic representation is shown in Fig.~\ref{fig:transition}(b.). 

As can be seen in Eq.~\ref{eq:Chi_3_case}, $\chi^{(3)}$ diverges when a resonance is reached, the denominator goes to zero, and the optical response of the system is significantly enhanced. 
A main limitation of this approach is that it is only applicable in the perturbative limit. However, in order to make use of excitons, it can be important to achieve population inversion. Fortunately, this has been achieved in atomic systems~\cite{dudovichSimpleRouteStrongField2005, trallero-herreroStrongfieldAtomicPhase2006,amitayMultichannelSelectiveFemtosecond2008,trallero-herreroTransitionWeakStrongfield2007, clowStrongFieldMultiphoton2008}.

The excitonic energy structure of mithrene around the gamma point is depicted in Fig.~\ref{fig:transition}(a.)~\cite{yaoStronglyQuantumConfinedBlueEmitting2021}. Bright states ($\mathrm{E_{C,A_{1s}}}$, s-symmetry), i.e. states that can be reached via one photon absorption, are colored blue, and dark states that require the absorption of two photons are colored red ($\mathrm{E_{C,A_{2p}},E_{C,B_{2p}}}$, p-symmetry). The emission (dashed arrows) from these states follow the same  selection rules, and thus the detection of dark states requires two photons. 
For excitation with $800~\mathrm{nm}$ light ($1.55~\mathrm{eV}$), there is no single photon excitation possible, and only TPA must be considered. Additionally, we treat the coupling between excited states as negligible and only allow for direct coupling between the ground state and the $\mathrm{A_{2p}}$ and $\mathrm{B_{2p}}$ exciton state.

This results in a modified version of the three-level Hamiltonian as derived before for the atomic case~\cite{trallero-herreroCoherentControlStrong2005,trallero-herreroTransitionWeakStrongfield2007}.

\vspace{3mm}
        \begin{equation}
            \centering
            \mathrm{\hat{\mathbf{H}}(t)=\left(
            \begin{array}{ccc}
                \omega _{g}^{(s)}(t) & \chi_\mathrm{A}(t)e^{i (\Delta_\mathrm{A} t-\varphi (t))} & \chi_\mathrm{B}(t)e^{i (\Delta_\mathrm{B} t- \varphi (t))}\\
                \chi_\mathrm{A} (t)e^{-i (\Delta_\mathrm{A} t-\varphi (t))} & \omega _{\mathrm{A}}^{(s)}(t) & 0\\
                \chi_\mathrm{B} (t)e^{-i (\Delta_\mathrm{B} t-\varphi (t))} & 0 & \omega _\mathrm{B}^{(s)}(t)
            \end{array}
            \right)     ,}
            \label{eq:3levelH}
        \end{equation}
\vspace{3mm}

\noindent where the diagonal terms $\mathrm{\omega_{g,A,B}}$ are the time-varying Dynamic Stark Shifts (DSS) for the ground and exciton states ($\mathrm{g}$, $\mathrm{A_{2p}/B_{2p}}$). $\mathrm{\chi_{A,B}}$ and $\mathrm{\Delta_{A,B}}$ are the coupling and detuning between the ground and $\mathrm{A_{2p}/B_{2p}}$ states, respectively. In the expression, $\mathrm{\omega_{g,A,B}^{(s)}}$ $\mathrm{\chi_{A,B} \propto I(t)\propto |E(t)|^2}$.

From the equation, it can be seen that in addition to the intensity and the center frequency, the phase can also be used as a control knob. The coupling of the states via $\mathrm{\chi_{A,B}}$ in off diagonal terms has very pronounced implications for any optical process, resonant or not. 
Unlike atoms, the width of dark exciton resonances is quite large. To compensate for this, we conducted multiple simulations with varying resonance energy. Further information regarding the simulations as well as the assumed parameters can be found in supplementary section 1.3. 

\subsection*{Experimental}

\begin{figure}
    \centering
    \includegraphics[width=0.95\linewidth]{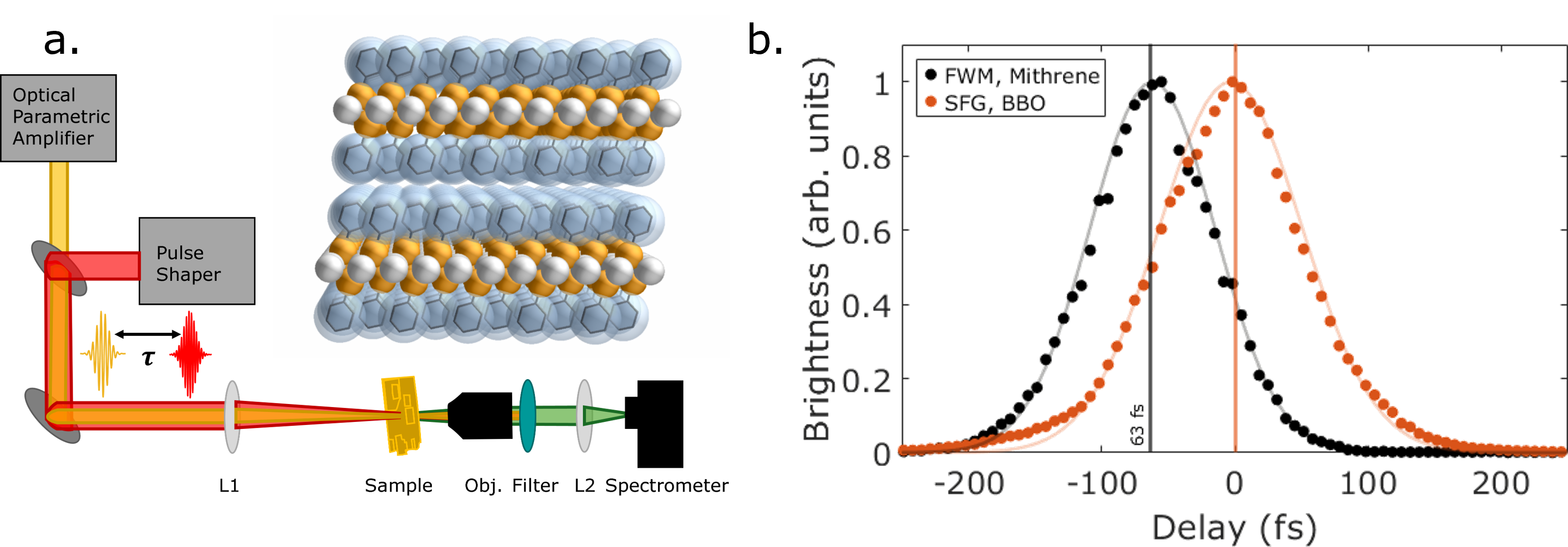}
    \caption{\textbf{(a.)} General experimental apparatus design used for all experiments. Pump pulses are are from a Ti:Sapphire laser and split between an OPA and pulse shaper. The resulting pulses are recombined and focused onto a sample of mithrene before the resulting emission is collected. \textbf{Inset.} Cartoon crystal structure of mithrene. Silver ball - Ag, Orange ball - Se, Blue hexagon/ball - phenyl ring ($\mathrm{C_6H_5}$)  \textbf{(b.)} Comparison of the FWM (black) brightness response in mithrene to the SFG (orange) brightness in BBO, both as a function of optical path length delay.}
    \label{fig:experiment_buildup}
\end{figure}

We utilize the technique of four wave mixing (FWM)~\cite{bauerExcitonicResonancesControl2022} to exploit the sensitivity of $\mathrm{\chi^{(3)}}$ to dark excitons. As already discussed,  FWM (through $\mathrm{\chi^{(3)}}$) increases dramatically at a resonance, allowing it to ``brighten'' dark states, as it provides an optical signal that can be readily measured via a spectrometer.

A sketch of the experimental setup is shown in Fig.~\ref{fig:experiment_buildup}(a.). The detected light follows the proposed photon scheme for the FWM depicted in Fig.~\ref{fig:transition} (two photon transitions).
The pump ($\mathrm{\lambda_{\mathrm{center}} =  795~nm}$, $40~\mathrm{fs}$) were tailored using an acusto-optical modulator (AOM) based
pulse shaper in a 4f geometry, where an arbitrary phase and amplitude can be applied
to each spectral component~\cite{warren_coherent_1993}, effectively controlling $\mathrm{\varphi(t)}$ and $\mathrm{\Delta_{A,B}}$  in Eq.~\ref{eq:3levelH}. The IR probe (yellow in Fig.~\ref{fig:transition}) was generated using an optical parametric amplifier (OPA) at $0.83~\mathrm{eV}$ ($1500~\mathrm{nm}$).
After the pump and probe pulses are recombined into a collinear geometry, both
pulses are focused onto the sample for various time delays, $\tau$. A more detailed description can be found in the Methods section.
By tuning the pump pulse wavelength, the FWM process can be resonant with either the $\mathrm{A_{2p}}$ or the $\mathrm{B_{2p}}$ state.  
The $\mathrm{A_{2p}}$ state for instance, requires a center wavelength of $816~\mathrm{nm}$, whereas the $\mathrm{B_{2p}}$ is in resonance with TPA of $787~\mathrm{nm}$ light. That being said, the difference is small enough such that both states can be accessed by a single $\sim20~\mathrm{fs}$ laser pulse centered at $800~\mathrm{nm}$.

\begin{figure}
    \centering
    \includegraphics[width=\linewidth]{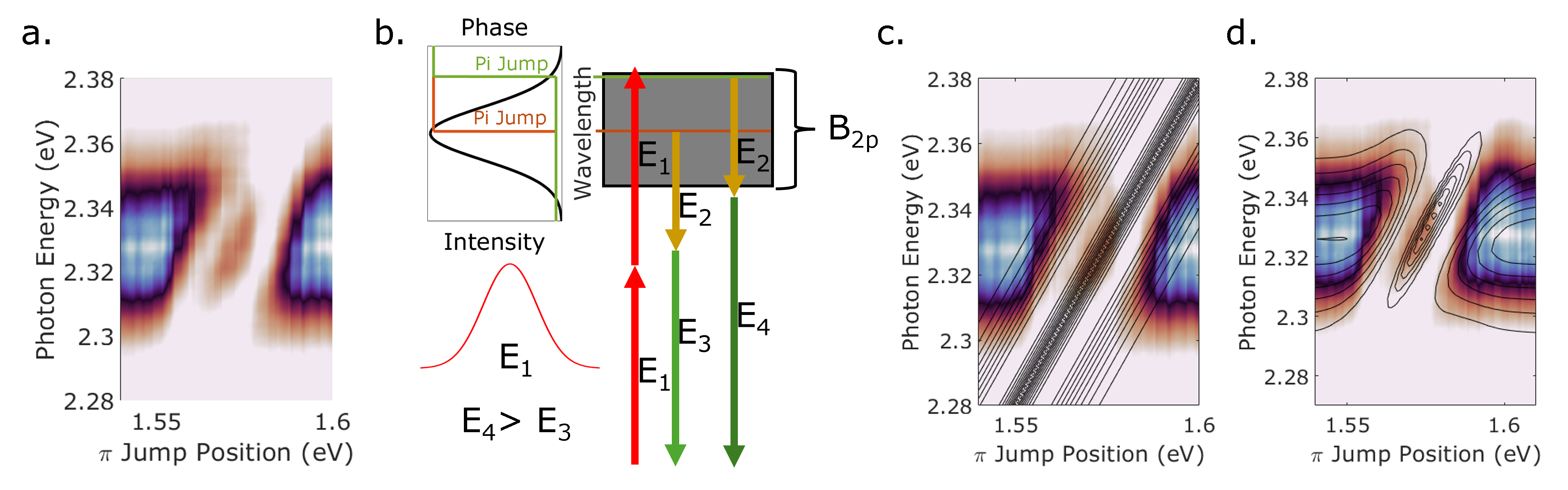}
    \caption{\textbf{(a.)} Two-dimensional heat map of the FWM emission as a function of the $\pi$ phase jump position. Constructive interference as a result of the $\mathrm{B_{2p}}$ resonance can be seen at $\sim1.575~\mathrm{eV}$. \textbf{(b.)} Illustration of the phase delay scanning through the pump spectrum, leading to constructive interference for the entire width of the $\mathrm{B_{2p}}$ resonance. \textbf{(c.)} $\mathrm{B_{2p}}$ state population resulting from second-order perturbative treatment of the Hamiltonian, Eq.~\ref{eq:3levelH}, overlayed over the heat map from (a.). \textbf{(d.)} Best theoretical agreement from a multidimensional scan of the integrated TDSE using the Hamiltonian from Eq.~\ref{eq:3levelH} compared to the heat map from (a.). Extracted material properties that fit the best for $\mathrm{I_0}=\mathrm{2.88 \cdot 10^{15}~W/m^2}$: $\mathrm{FWHM_{pump}}=\mathrm{12~nm}$, $\mathrm{\lambda_{pump}}=\mathrm{786~nm}$,     $\mathrm{E_{B_{2p}}}=\mathrm{3.16~eV}$, $\mathrm{\omega_{B_{2p}}}=\mathrm{-54.9~Trad/s}$,    $\mathrm{\chi_{B_{2p}}}=\mathrm{1.13\cdot 10^{-2}~Trad/s}$}
    \label{fig:pi_jump}
\end{figure}

To demonstrate control beyond perturbative treatment, we perform experiments using one of the simplest phase modulation cases, a spectral $\pi$ phase flip scanned
across the spectrum, 

\vspace{3mm}

\begin{equation}
    \mathrm{\phi(\lambda)=}
    \begin{cases}
    \mathrm{0,~\lambda < \lambda_\pi}\\
    \mathrm{\pi,~\lambda \geq \lambda_\pi}\\
    \end{cases}
    \label{eq:pi_jump}
\end{equation}

\vspace{3mm}

The interpretation of this spectral phase is rather simple. For a system with a resonance, TPA is maximum when $\mathrm{\lambda_{\pi}}$ is at the resonance of the two photon frequency~\cite{meshulachCoherentQuantumControl1998,trallero-herreroTransitionWeakStrongfield2007,amitayMultichannelSelectiveFemtosecond2008}. In the strong field limit, when the excited state population is $>> 0$, the DSS also plays a role in the structure of the excitation. In cases where there is no resonance at the two-phton frequency, the $\pi$-flip simply follows the peak of the spectrum. In other words, for no TPA resonance, $\mathrm{\lambda_\pi = \lambda_L}$~\cite{meshulachCoherentQuantumControl1998}.
Temporally, this translates into different times for the buildup for resonant systems compared to non-resonant or parametric ones. In Fig.~\ref{fig:experiment_buildup}(b.), comparing the pump-probe scans of the FWM in mithrene to similar scans in a BBO crystal, it becomes clear that the FWM from mithrene is not parametric.

Mithrene does not allow second-order optical processes due to its centrosymmetric nature~\cite{cuthbertSynthesisStructuralCharacterization2002,schriberMithreneSelfAssemblingRobustly2018}. Thus, any observed structure, in addition to the deconstructive interference, must originate from a resonant state.

The results of a $\pi$-jump scan in mithrene are shown in Fig.~\ref{fig:pi_jump}(a.) where the emission energy is plotted as a function of the $\pi$-jump position. 
The destructive interference can be observed between $1.55~\mathrm{eV}$ and $1.6~\mathrm{eV}$ with a width of constructive emission at $1.575~\mathrm{eV}$.
This pump energy corresponds to the $\mathrm{B_{2p}}$ dark exciton resonance ($2~\mathrm{x}~1.575~\mathrm{eV}$) of mithrene. 
In addition to showing control of the state population, this simple scan furthermore reveals that the resonance has a finite width of at least $>0.01~\mathrm{eV}$. A schematic representation of this can be seen in Fig.~\ref{fig:pi_jump}(b.) where different energies of the $\mathrm{B_{2p}}$ resonance can be accessed by scanning the phase delay through the pump spectrum. This observation is in line with the assumption that the absorption of two pump photons can be used for exclusive state access. TPA followed by interaction with the probe pulse transforms the dark, second-order process into an optically allowed third-order FWM process, which can be easily observed, yet still carries the information of the two photon transition.

Although the perturbation theory calculations (Fig.~\ref{fig:pi_jump}(c.)) demonstrate the general shape, they fail to capture the specific features of the FWM signal.
For visibility, simulation results are overlaid with the experimental results of panel (a.). One clear conclusion from perturbation theory is that (as mentioned before), even parametric processes have regions of destructive and constructive interference. These regions are perfectly parallel and follow the center wavelength of the two participating photons.
To better match the experimental results, we move to a simulation that utilizes strong field excitation. This was done by integrating the time-dependent Schr\"odinger equation using the Hamiltonian outlined in Eq.~\ref{eq:3levelH}.
Fig.~\ref{fig:pi_jump}(d.) shows these simulation results overlaid with experimental data. As most material constants of mithrene are unknown, multiple simulations were run and the closest match to the experiment is depicted. 
Although exact values for the state coupling and DSS cannot be determined by this calculation, it was concluded that the DSS moves the $\mathrm{B_{2p}}$ state closer to the ground state, resulting in a slight upward bending of the FWM signal around $1.57~\mathrm{eV}$ ($\pi$-jump position) and $2.36~\mathrm{eV}$ (FWM signal energy). Further details about the simulations and their limitations can be found in supplemental section 1.3.

\begin{figure}
    \centering
    \includegraphics[width=0.95\linewidth]{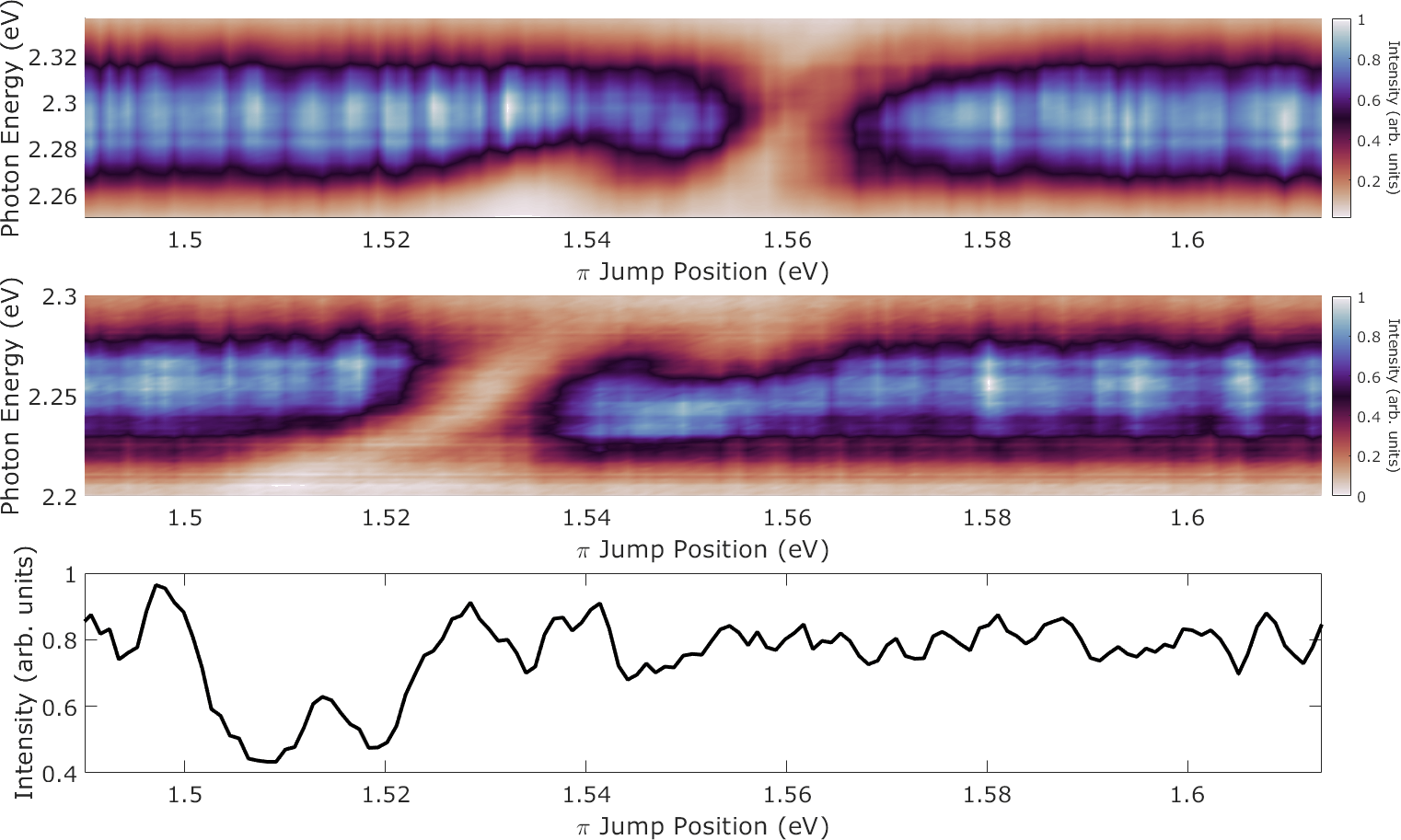}
    \caption{\textbf{(a.)} $\mathrm{\lambda_{center}}$ = $1.56~\mathrm{eV}$ \textbf{(b.)}. $\mathrm{\lambda_center = 1.53~eV}$. \textbf{(c.)} Lineout of the emission centered at $2.19~\mathrm{eV}$.} 
    \label{fig:far_off}
\end{figure}

In addition to control of $\mathrm{\varphi(t)}$ of the excitation light, the pulse shaper also allows for changing the center wavelength of the light and thus the detuning relative to the resonances, i.e. 
$\mathrm{\Delta_{A,B}}$ from Eq.~\ref{eq:3levelH}. One would expect that the shape of the FWM will change dramatically once the detuning is larger than the bandwidth of the resonance. 
Figure~\ref{fig:far_off} depicts $\pi$-phase-jump scans for two different center wavelengths of the pump pulse, Fig.~\ref{fig:far_off}(a.) $1.56~\mathrm{eV}$ and (b.) $1.53~\mathrm{eV}$, both significantly detuned from any resonance. As can be seen, the destructive interference is still present yet no constructive interference is observed, corroborating that the constructive interference requires an allowed two photon transition. Looking closer at the shape of the FWM signal in both Fig.~\ref{fig:far_off}(a.) and (b.), the destructive interference around the center wavelength is observed with additional pronounced regions of destructive interference visible at energies less than $1.54~\mathrm{eV}$ and above $1.56~\mathrm{eV}$, but below the $\mathrm{B_{2p}}$ resonance. Figure~\ref{fig:far_off}(b.) especially has a very pronounced dip on the red side of the excitation is visible. We attribute the emerging dip in the FWM signal to the edge of the $\mathrm{A_{2p}}$ resonance, which is expected at 1.52 eV. Looking at a lineout (Fig.~\ref{fig:far_off}(c.)) below the prominent FWM signal, at the very edge of our available bandwidth, centered at $2.19~\mathrm{eV}$ and averaged across neighboring pixels, hints of a feature that corresponds to a resonance are revealed. The location is approximately at the energy of the $\mathrm{A_{2p}}$ resonance. This suggests that with an additional broadening stage of our pump pulse, we could access both $\mathrm{A_{2p}}$ and $\mathrm{B_{2p}}$ individually using only a pulse shaper.

\section*{Discussion}

With this work, we have shown coherent quantum control of the population of the dark excitonic states of mithrene in the non-perturbative limit. The developed three level Hamiltonian promises deeper insight into how to manipulate the states in order to use them for quantum control. One way forward is to identify conditions as well as material configurations that allow for different quantum operations on one (or both) states by manipulating the pulse shape (e.g., intensity, phase, center wavelength, pulse duration, etc.). MOChas provide the perfect platform for this due to their basically unlimited tunability. Modification of any of the components (metal, chalcogen, or ligand) have been shown to impact not only the material’s dimensionality, but also its band structure and emissive properties~\cite{abdallahLuminescentandSustainable2023,fanNucleophilicDisplacementReactions2024,willsonSupramolecularSupport2025}. Recent reports~\cite{tomoakiSyntheticBandGap2025,paritmongkolSizeQualityEnhancement2021}, have enabled the transition from powders to single crystals large enough for nonlinear and ultrafast optical studies.
In a more general sense, using two-photon transitions (independent on the system) will open the door to a new path for quantum information and processing.

\subsection*{Application to Quantum Processing}
Indeed, one of the most exciting prospects of this work is its direct application to quantum processing and quantum information \cite{nielsenQuantumComputationQuantum2011}. Some theoretical approaches have been proposed for qubit realizations with excitons \cite{ghoshQuantumComputingExcitonpolariton2020,castellanosDesignMolecularExcitonic2020,harankahageQuantumComputingExciton2021}, but to the best of our knowledge, such proposals do not include the use of multiple dark excitons; the closest is the utilization of the spin states of a single dark exciton \cite{schwartz2015deterministic}. Here we show that when using multiphoton transitions, the inclusion of the laser field phase as a continuous variable can be used for the generation of generalized quantum gates.

Given the Hamiltonian, Eq.~\ref{eq:3levelH}, and assuming a narrow bandwidth pump, the system can be reduced to a two-level system that includes $\vert \mathrm{g}\rangle$ and $\vert \mathrm{B}\rangle$ states (see Supplement Sect. 2). Applying the unitary transformation 

\begin{equation*}
\mathrm{U(t)=\exp\!\left\{-i\vartheta(t)\,\vert B\rangle\langle B\vert\right\}}
=
\begin{pmatrix}
1 & 0\\
0 & \mathrm{e^{-i\vartheta(t)}}
\end{pmatrix},
\end{equation*}
one arrives at an effective two-level Hamiltonian that depends on the two photon coupling $\mathrm{\chi(t)}$ and the strong field detuning parameter $\mathrm{\delta(t)=\Delta_B-\dot\phi(t)-\Delta\omega^{(s)}(t)}$. 

\paragraph*{$\mathbf{R_x(\frac{\pi}{2})}$ Gate}
From the definition of $\mathrm{\delta(t)}$, it can be seen that the on-resonant tracking condition is given by $\mathrm{\delta(t)=0}$, which can be obtained by manipulating the laser phase $\mathrm{\phi(t)}$. Under this condition, we obtain a $\mathrm{\pi/2}$ Hamiltonian, $\mathrm{H_{\text{rel}}(t)=\chi(t)\sigma_x}$, and gate
\begin{equation*}
\mathrm{U_x(\theta)=\exp\!\left(-i\frac{\theta}{2}\sigma_x\right),}   
\end{equation*}
with the pulse area given by $\mathrm{\theta=2\int_{\text{pulse}}\chi(t)\,dt}$

\paragraph*{Hadamard Gate}
Similarly, it is possible to find a pulse shape for a Hadamard gate. In this case, the necessary condition is 
$\mathrm{\vert\delta(t)\vert \gg \vert2\chi(t)\vert}$ with a generalized Rabi frequency, $\mathrm{\Omega(t)=\sqrt{4\chi^2(t)+\delta^2(t)}}$. The corresponding Hamiltonian is then $\mathrm{H_{\text{phys}}(t)
=\frac{1}{2} \left(2\chi(t)\sigma_x + \delta(t)\sigma_z \right)}$ with a unitary gate,
\begin{equation*}
\mathrm{U_z(\varphi)=
\exp\!\left(-i\frac{\varphi}{2}\sigma_z\right),}
\end{equation*}
$\mathrm{\varphi
=\int \left(\delta(t) + \frac{4\chi^2(t)}{\delta(t)} \right) dt}$ up to a second order correction.

Therefore,

\begin{equation*}
\mathrm{U_H
\sim
U_z\!\left(\frac{\pi}{2}\right)
U_x\!\left(\frac{\pi}{2}\right)
U_z\!\left(\frac{\pi}{2}\right),}
\end{equation*}

From the above equations, it is clear that other gates are possible. Furthermore, as pointed out in \cite{trallero-herreroStrongfieldAtomicPhase2006}, the number of solutions for a multiphoton $\pi$ pulse is infinite, because of the presence of the phase. Therefore, there are many gate solutions that involve composite-pulses (two or more pulses) that are more robust to central phase detunning.
Finally, for mithrene specifically, using a slightly broader pump pulse in the future will give access to both the $\mathrm{A_{2p}}$ and the $\mathrm{B_{2p}}$ states at the same time yet independently. This will allow for a qubit defined in the $\mathrm{\{\vert A\rangle,~\vert B\rangle\}}$ space with an even higher degree of control. A forthcoming publication will explore a set of such quantum gates.

Additionally, the 4f pulse shaper is not limited to a single excitonic site, but allows for a 1D array of excitons by simply driving the pulse shaper with a superposition of rf-driving frequencies. While a theory for the interaction of an array of dark excitons is still not complete, we can envision interactions similar to those present in polariton condensates~\cite{topferEngineeringSpatialCoherence2021}. These two facts can be combined to create a multi-qubit processing unit in MOChas, or any strongly two-photon coupled material.

\section*{Materials and Methods}

\subsection*{Sample Preparation}

Silver nitrate ($\geq 99.0~\%$), diphenyl diselenide ($98~\%$), octylamine ($99~\%$), and toluene ($\geq99.9~\%$) were used as received from Sigma-Aldrich. Twenty-mL glass scintillation vials were purchased from Thermo Fisher Scientific. 

Based on previously reported procedure \cite{paritmongkolSizeQualityEnhancement2021}, mithrene single crystals were produced in a single-phase reaction in which $10~\mathrm{mL}$ of a $3~\mathrm{mM}$ solution of silver nitrate in octylamine was mixed with $10~\mathrm{mL}$ of $3~\mathrm{mM}$ diphenyl diselenide in toluene in a glass vial. The vial was set at room temperature for $3-5$ days until the crystals had grown to the desired size.

After synthesis, the thin flakes are transferred and attached to the sample holder. The utilized holder has a through hole in the center, which allows for transmission measurements. The flakes themselves are positioned such that the thinnest parts, i.e. edges, are positioned over this region.

\subsection*{Laser System}

The employed laser source is a Ti:Sapphire based system (Continuum USA) with a central wavelength of $\sim800~\mathrm{nm}$ ($1.55~\mathrm{eV}$), $1~\mathrm{kHz}$ repetition rate, $\sim12~\mathrm{mJ}$ pulse energy, and a pulse duration of $20~\mathrm{fs}-40~\mathrm{fs}$. $1~\mathrm{mJ}$ of the laser output was split $70:30$ between an Optical Parametric Amplifier (Spectra-Physics OPA-800CF) and a pulse shaper, respectively. The OPA was tuned to $\sim1500~\mathrm{nm}$ ($0.827~\mathrm{eV}$), $50~\mathrm{fs}-70~\mathrm{fs}$ duration. The pulse shaper was configured in a $4f$ geometry using an acousto-optical modulator (AOM, Brimrose, $\mathrm{TeO_2}$ longitudinal mode, $150~\mathrm{MHz}$) in the Fourier plane and gold-coated ruled diffraction gratings (Richardson Grating Laboratory, $671.64~\mathrm{lines/mm}$). The acoustic waveform was synthesized with an arbitrary waveform generator (Chase Scientific, DAx11500z-LAN). The pump and probe pulses were then brought collinear and focused on the mithrene sample or a BBO crystal before the emission spectra were collected using an objective (Thorlabs, LMU-15X-NUV) and focused into a fiber coupled spectrometer (Ocean Optics USB2000). The pump and probe pulses were filtered with a bandpass filter (Thorlabs, FGB39) that only transmits the FWM signal.

\subsection*{Strong Field Simulations and Quantum Gates Derivations}

The parameter space that has been investigated can be found in the supplemental section 1.3. The derivations for the quantum gates can be found in supplemental section 2.

\subsection*{Comparison of Theory to Experiment}

The comparison of simulation results to experimental results is based on the mean-root-square (RMS) of the difference of the normalized  emission as a function of $\pi$-jump position to the normalized theoretical population. The normalization of the population was conducted by setting the integral of the region of interest to 1. The region of interest was set to the $\pi$-jump range of $1.55-1.59~\mathrm{eV}$. Due to the significant bandwidth of the probe this parameter was optimized within the range of $0.81-0.83~\mathrm{eV}$. A zoomed in histogram of the best RMS values is added in Fig.~S1.  

\subsection*{Acknowledgments}
TS and CLM were partially supported by US Department of Energy,
Office of Science, Chemical Sciences, Geosciences, \& Biosciences Division grant DE-SC0024508. Instrumentation was partially supported by the Office of the Vice President for Research at the University of Connecticut. MA, MCW, and JNH were supported by the US Department of Energy Integrated Computational and Data Infrastructure for Scientific Discovery grant DE-SC0022215. MCW gratefully acknowledges the National Science Foundation Graduate Research Fellowship Program under grant no. DGE 2136520. 

\subsection*{Conflicts of Interest}
C.T-H is the founder and sole owner of QueHot, LLC, a company invested in developing quantum architecture. All other co-authors declare no conflict of interest.



\section{Supplement}

\section{Theory}
    \subsection{Perturbative Regime}
        For a two-photon transition with the consideration of no intermediate states, ie. the detuning of the intermediate states is sufficiently large that they are not populated, we can use a basic two-level Hamiltonian. In a similar second-order perturbation theory calculation as Ref.~\cite{trallero-herreroTransitionWeakStrongfield2007}, the excited state amplitude of the $\mathrm{A_{2p}/B_{2p}}$ resonances yields: 

        \begin{multline}
	       a_{A_{2p}/B_{2p}} = -\frac{1}{\hbar^2} \sum_{m} \mu_{A_{2p}/B_{2p},m}\mu_{m,g} \int_{-\infty}^{t_2} \int_{-\infty}^{t_1} \mathcal{E}\left(t_1\right)\mathcal{E}\left(t_2\right)\\
           exp\left(-i \Delta_{A_{2p}/B_{2p},m}t_1\right) exp\left(-i \Delta_{m,g}t_2\right) \,dt_1 \,dt_2,
           \label{eq:A2pB2p_amplitude}
        \end{multline}

        where $\omega_0$ is the laser frequency, $\Delta_{ij}$ is the detuning between the transition energy and the laser frequency. Taking the inner integral with the assumptions that there is a sufficiently large intermediate detuning, $\exp(-i\Delta_{m,g}t_2)$ varies faster than $\mathcal{E}(t_2)$, and taking the limit of $t \rightarrow\infty$, the probability of the absorption into either dark exciton state becomes~\cite{trallero-herreroTransitionWeakStrongfield2007}

        \begin{equation}
	        P_{g\rightarrow A_{2p}/B_{2p}} = \left| \frac{\left<A_{2p}/B_{2p}|\mu^2|g\right>}{\bar{\omega}\hbar^2} \right|^2 \left| \int_{-\infty}^{\infty} \mathcal{E}^2 \left(t\right) exp\left(-i\Delta t \right) \,dt \right|.
            \label{eq:TPA_prob_time}
        \end{equation}

        The term $\hbar\bar{\omega}$ is the average resonance energy of the parity-allowed, intermediate states. Finally, effective two-photon coupling is represented as $\left<A_{2p}/B_{2p}|\mu^2|g\right>$. In the frequency domain, the same TPA probability can be written as
        
        \begin{multline}
	           P_{g\rightarrow A_{2p}/B_{2p}} \propto \left\lvert \int_{-\infty}^{\infty} \Tilde{\mathcal{E}}\left( \omega_{A_{2p}/B_{2p},g}/2 + \Omega \right) \Tilde{ \mathcal{E}}\left( \omega_{A_{2p}/B_{2p},g}/2 - \Omega \right) \,d \Omega \right\rvert^2 \\
	           \propto \left\lvert A\left( \omega_{A_{2p}/B_{2p},g}/2 + \Omega \right) A\left( \omega_{A_{2p}/B_{2p},g}/2 - \Omega \right) \times e^{\Phi\left( \omega_{A_{2p}/B_{2p},g}/2 + \Omega \right) \Phi\left( \omega_{A_{2p}/B_{2p},g}/2 - \Omega \right)} \,d \Omega \right\rvert ^2, 
            \label{eq:TPA_prob_freq}
        \end{multline}

        where $\mathcal{E}(\omega)$ is the Fourier transform of the time domain electric field, $\mathrm{A(\omega)}$ is the spectral amplitude, and $\Phi(\omega)$ is the spectral phase.

        In a perturbative regime, the third order non-linear susceptibility, $\chi^{\left(3\right)}$, is highly sensitive to intermediate resonance states. This is seen when the $\omega_{ng}-\omega_q-\omega_p$ term in Eq.~\ref{eq:Chi_3} becomes resonant or nearly resonant with the state at $\omega_{ng}$~\cite{boydNonlinearOptics2020}.

        \begin{equation}
                \chi^{\left(3\right)}_{kjih} \left(\omega_{\sigma}, \omega_r, \omega_q, \omega_p \right) \propto \frac{\boldsymbol{\mu}^j_{g\nu}\boldsymbol{\mu}^k_{\nu n}\boldsymbol{\mu}^i_{mg}\boldsymbol{\mu}^h_{\nu n}}{\left(\omega^{\ast}_{\nu g} + \omega_r\right)\left(\omega_{ng}-\omega_q-\omega_p\right)\left(\omega_{mg}-\omega_p\right)}
            \label{eq:Chi_3}
        \end{equation}

        In the case of \textit{mithrene}, the transition $\omega_{ng}$ represents the two photon transition from the valance band to the dark exciton states $\mathrm{A_{2p}}$ and $B_{2p}$. This makes for a near-degenerate Four Wave Mixing process $\left(2\omega_{pump} - \omega_{probe} \right)$ ideal as $\omega_{pump}$ can be tuned to be resonant with dark excitons.

    \subsection{Non-perturbative Regime}

        The induced polarization of the two-level system due to the pump field can be described in a similar fashion as Trallero et. al.~\cite{trallero-herreroUnderstandingStrongfieldCoherent2006}:

        \begin{multline}
            \frac{\partial}{\partial t}\mathcal{E}_L(z,t) = \\ i\hbar\frac{N\omega_0}{2}
            \left[Q_{11}(z,t)\omega_g^{(s)}\mathcal{E}_L(z,t) + Q_{22}(z,t)\omega_{A_{2p}/B_{2p}}^{(s)}\mathcal{E}_L(z,t) + 2Q_{21}(z,t)\chi e^{-i\Delta t}\mathcal{E}_L^*(z,t)\right],
            \label{eq:pump_field_description}
        \end{multline}

        where $\mathcal{E}_L(z,t)$ is the shaped or unshaped field, $Q_{ij}$ are the slow varying density matrix elements for the ground and $A_{2p}/B_{2p}$ states. Given the dark nature of the exciton states, emission at frequency $\omega_{A_{2p}/B_{2p}}$ is non-existent. However, coupling the state with near-degenerate FWM allows for observation of the state. The effective third order susceptibility for this process can be described by:

        \begin{multline}
            \chi^{(3)}_{eff}[\omega_{FWM} = 2\omega_{pump}-\omega_{probe}] = \\
            i\hbar\frac{N\omega_0}{6\epsilon_0\mathcal{E}_L^2\mathcal{E}_1^*} [Q_{11}(z,t)\omega_g^{(s)}\mathcal{E}_L(z,t) + Q_{22}(z,t)\omega_{A_{2p}/B_{2p}}^{(s)}\mathcal{E}_L(z,t) + 2Q_{21}(z,t)\chi e^{-i\Delta t}\mathcal{E}_L^*(z,t)].
            \label{eq:nonperturb_chi3}
        \end{multline}

        Given the assumption the state couples with the virtual state created by the FWM process, we effectively have a three level problem where the evolution of the the population between the dark state the virtual state can be described in a similar fashion to~\cite{trallero-herreroUnderstandingStrongfieldCoherent2006}:

        \begin{multline}
            \frac{\partial}{\partial t} Q_{3,2} = i [\omega_{e}^{\left(s\right)}\left(z,t\right)Q_{3,2} + \chi\left(z,t\right)e^{i\Delta t}Q_{3,1} + \\
            \frac{\mu_{3,2}}{2 \hbar}\mathcal{E}_{1}^{\ast}\left(z,t\right) e^{i\Delta_{2,3} t}\left(Q_{22}-Q_{33}\right) + \frac{\mu_{3,1}}{2 \hbar}\mathcal{E}_{2}\left(z,t\right) e^{i\Delta_{3,1} t} Q^{\ast}_{21}]
            \label{eq:Dark2FWM_evolution}
        \end{multline}

    \subsection{Simulation parameters}

        As most material constants of \textit{mithrene} are unknown we start with the well studied Sodium as the baseline. This leads to the following values for the stark shift $\omega$ as well as the coupling, $\chi$, between the ground and resonant state~\cite{trallero-herreroCoherentControlStrong2005}: 


\begin{table}[h]
    \centering
    \caption{Material properties of atomic Sodium for Intenisty $\mathrm{I_0}$ \cite{trallero-herreroCoherentControlStrong2005}.}
    \begin{tabular}{cc}
        \hline
        \hline
        \\
        $\mathrm{I_0}$ & $\mathrm{2.883 \cdot 10^{15}~W/m^2}$  \\
        $\mathrm{\chi_{Na}(I_0)}$ & $\mathrm{1.0234 \cdot 10^{-2}~Trad/s}$  \\
        $\mathrm{\omega_{A_{2p}/B_{2p}}(I_0)}$ & $\mathrm{18.3~Trad/s}$  \\
        \\
        \hline
        \hline
    \end{tabular}
    	\label{table:Ham_parms_start}
\end{table}

        Additionally we assume $\omega_{ground~state}=0$ and the excitonic states to have a bandwidth $\Delta E_{A2p,B2p} = 0.2~eV$. To find the closest match to the experimental data, which has mainly access to the B2p exciton, the following parameter space has been searched:


        \begin{table}
    \centering
    \caption{Explored parameter space in the simulations.}
    \begin{tabular}{ccc}
        \hline
        Parameter & Range & Resolution  \\
        \hline
        \hline
        \\
        Intensity & $\mathrm{(1.4415~-~7.2075)\cdot10^{15}~W/m^2}$ & $\mathrm{1.4415\cdot10^{15}~W/m^2}$ \\
        $\mathrm{FWHM_{pump}}$ & $\mathrm{(12~-~16)~nm}$ & $\mathrm{1~nm}$ \\
        $\mathrm{\lambda_{pump}}$ & $\mathrm{(784~-~789)~nm}$ & $\mathrm{1~nm}$  \\
        $\mathrm{E_{B_{2p}}}$ & $\mathrm{(3.14~-~3.18)~eV}$ & $\mathrm{0.01~eV}$  \\
        $\mathrm{\omega_{B_{2p}}}$ & $\mathrm{(-54.9~-~+54.9)~Trad/s}$ & $\mathrm{18.3~Trad/s}$ \\
        $\mathrm{\chi_{B_{2p}}}$ & $\mathrm{(1.0234\cdot10^{-3}~-~1.63744\cdot10^{-2})~Trad/s}$ & $\mathrm{0.5117\cdot10^{-2}~Trad/s}$    \\
        \\
        \hline
        \hline
    \end{tabular}
    \label{table:Ham_parm_ranges}
\end{table}

\vspace{3mm}

        Finally the RMS to experiment was calculated within the $\pi$ jump range of 1.55-1.59~eV. Due to the significant bandwidth of the probe pulse we assumed the signal wavelength energy to be between 0.81-0.83~eV and only compared the lowest rms. 

\begin{figure}[h]
            \centering
            \includegraphics[width=0.7\linewidth]{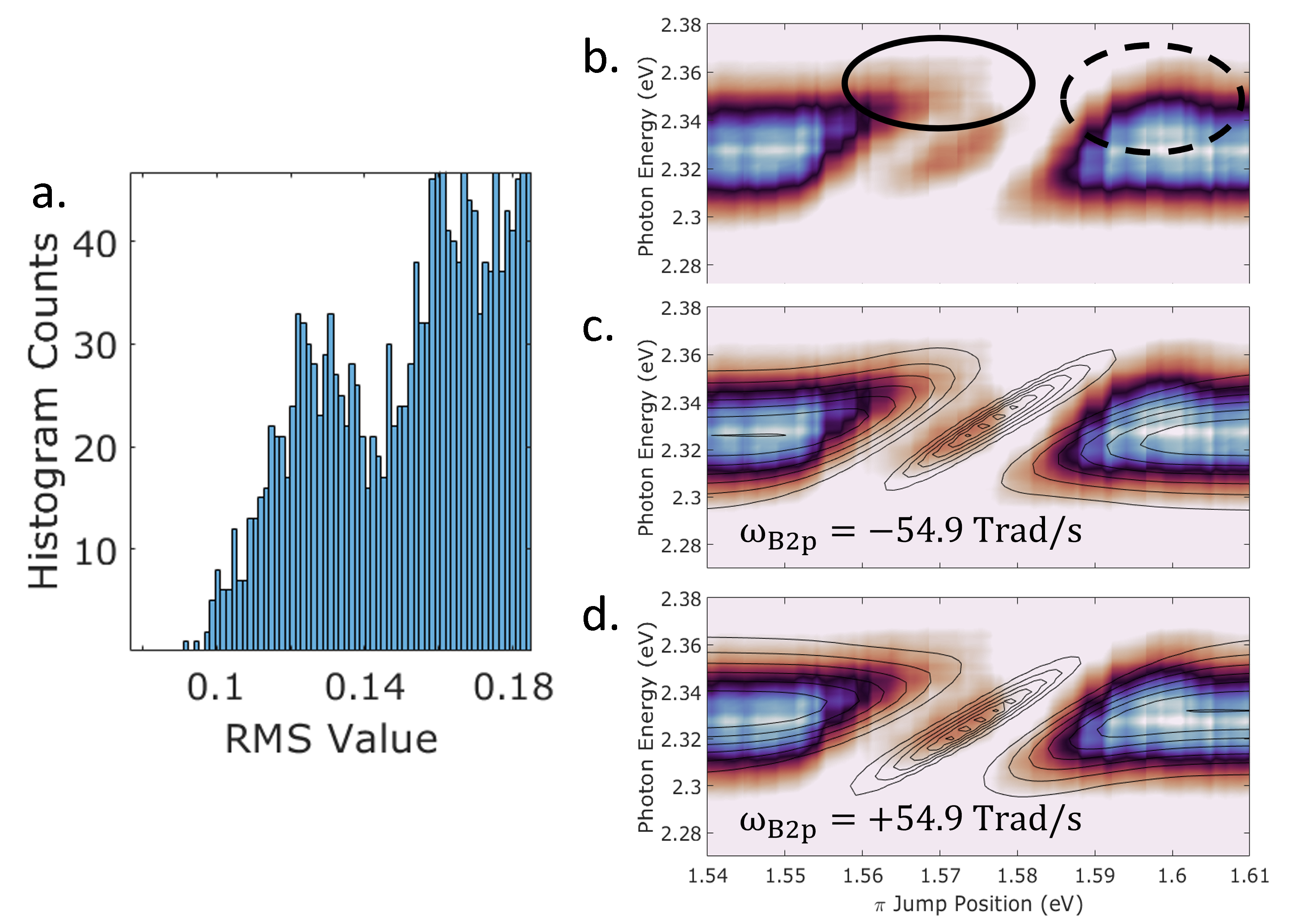}
            \caption{(a) Zoom in of the histogram for all simulated RMS values. (b) experimental $\pi$ jump data. Solid and dashed circle mark regions of interest, see text. (c)(d) Simulation result for an $\omega_{B2p}$ of -54.9~Trad/s (RMS of 0.092) and +54.9~Trad/s (RMS of 0.102) overlayed with the experimental data respectively. The rest of the material properties is kept constant.}
            \label{fig:theory_comp}
        \end{figure}

        Figure \ref{fig:theory_comp} (a) shows a zoom in on the best rms values. It is obvious that with the current simulations and the room temperature sample many simulation parameters result in similar results. However, the 5 best parameter sets have all a negative $\omega_{B2p}$ ranging from -36.6~Trad/s to -54.9 Trad/s, which results in a decrease of the energy difference between the ground state and the exciton state. On the right of fig. \ref{fig:theory_comp} the original experimental data (b) as well as -54.9 Trad/s (c) and +54.9 Trad/s (d) $\omega_{B2p}$ is depicted.
        We want to point towards two regions marked in (a). The solid area fits significantly better for negative $\omega_{B2p}$ as it captures the slight up-bending of the signal, whereas the dashed area is not captured with any simulation parameter. This points towards weaknesses in the simulation, most likely the finite width of the resonance as well as coupling to phonon states. Both effects will be weakened for lower temperatures. Additionally, it is rather striking that the regions of destructive interference are not symmetric around the constructive interference, which is in contrast to the perturbation theory.

        Regarding the coupling constant $\chi_{B2p}$ the best rms values were achieved at $1.12574\cdot 10^{-2}$~Trad/s .

\section{Derivation of universal quantum gates}
We work in the ordered basis $\{|g\rangle,|B\rangle\}$ and define
\[
\vartheta(t)\equiv \Delta_B t-\phi(t),
\qquad
\chi(t)\equiv \chi_B(t)\in\mathbb{R}.
\]
The lab-frame Hamiltonian is
\begin{equation}
H(t)=
\begin{pmatrix}
\omega_g^{(s)}(t) & \chi(t)e^{+i\vartheta(t)}\\
\chi(t)e^{-i\theta(t)} & \omega_B^{(s)}(t)
\end{pmatrix}.
\label{eq:H_lab_real}
\end{equation}

\subsubsection*{Rotating-Frame Transformation}

Choose
\begin{equation}
U(t)=\exp\!\left\{-i\theta(t)\,|B\rangle\langle B|\right\}
=
\begin{pmatrix}
1 & 0\\
0 & e^{-i\theta(t)}
\end{pmatrix},
\end{equation}

to yield the rotated Hamiltonian,
\begin{equation}
H'(t)=
\begin{pmatrix}
\omega_g^{(s)}(t) & \chi(t)\\
\chi(t) & \omega_B^{(s)}(t)-\Delta_B+\dot\phi(t)
\end{pmatrix}.
\label{eq:Hprime_real}
\end{equation}

Subtracting $\omega_g^{(s)}(t)\mathbb I$ and define the differential Stark shift
\[
\Delta\omega^{(s)}(t)\equiv \omega_B^{(s)}(t)-\omega_g^{(s)}(t).
\]
Then
\begin{equation}
H_{\text{rel}}(t)=
\begin{pmatrix}
0 & \chi(t)\\
\chi(t) &
\Delta\omega^{(s)}(t)-\Delta_B+\dot\phi(t)
\end{pmatrix}.
\end{equation}

Define the detuning
\begin{equation}
\delta(t)=\Delta_B-\dot\phi(t)-\Delta\omega^{(s)}(t),
\end{equation}
so that
\begin{equation}
H_{\text{rel}}(t)=
\begin{pmatrix}
0 & \chi(t)\\
\chi(t) & -\delta(t)
\end{pmatrix}.
\end{equation}

\subsection*{Tracking Phase or Atomic Phase Matching Condition}

As shown in \cite{trallero-herreroTransitionWeakStrongfield2007,trallero-herreroCoherentControlStrong2005} to preserve the resonant condition ($\delta(t)=0$), the laser phase needs to evolve according to the DSS,

\begin{equation}
\dot\phi(t)=\Delta_B-\Delta\omega^{(s)}(t).
\end{equation}

\begin{equation}
\phi(t)=\phi_0+\Delta_B t-\int_0^t \Delta\omega^{(s)}(t')dt'.
\end{equation}

\subsection*{Single-Pulse Gate with Real $\chi(t)$}

Under resonance ($\delta=0$),
\begin{equation}
H_{\text{rel}}(t)=
\chi(t)\sigma_x.
\end{equation}
Define pulse area
\begin{equation}
\theta=2\int_{\text{pulse}}\chi(t)\,dt.
\end{equation}
The resulting unitary is
\begin{equation}
U_x(\theta)=\exp\!\left(-i\frac{\theta}{2}\sigma_x\right).
\end{equation}

For the ordered basis set, $\{ |g>, |B>\}$ where $|g>=0$, $|B>=1$,

\[
\sigma_x =
\begin{pmatrix}
0 & 1 \\
1 & 0
\end{pmatrix},
\qquad
\sigma_y =
\begin{pmatrix}
0 & -i \\
i & 0
\end{pmatrix},
\qquad
\sigma_z =
\begin{pmatrix}
1 & 0 \\
0 & -1
\end{pmatrix}.
\]

\subsection*{$\mathrm{R_x(\frac{\pi}{2})}$ Gate}

A $\pi/2$ rotation about $x$ $(\mathrm{R_x(\frac{\pi}{2})})$requires
\[
\theta=\frac{\pi}{2}.
\]
Thus
\begin{equation}
U_{X_{\pi/2}}=
\exp\!\left(-i\frac{\pi}{4}\sigma_x\right).
\end{equation}

\bibliography{sn-bibliography}

\end{document}